# Related or Unrelated Diversification: What is Smart Specialization?

Önder Nomaler and Bart Verspagen

UNU-MERIT

25 March 2023

**Abstract**:

In this paper, we investigate the nature of the density metric, which is employed in the literature on smart specialization and the product space. We find that although density is supposed to capture relatedness between a country's current specialization pattern and potential products that it may diversify into, density is also correlated strongly to the level of diversification of the country, and (less strongly) to the ubiquity of the product. Together, diversity and ubiquity capture 93% of the variance of density. We split density into a part that corresponds to related variety, and a part that does not (i.e., unrelated variety). In regressions for predicting gain or loss of specialization, both these parts are significant. The relative influence of related variety increases with the level of diversification of the country: only countries that are already diversified show a strong influence of related variety. In our empirical analysis, we put equal emphasis on gains and losses of specialization. Our data show that the specializations that were lost by a country often represented higher product complexity than the specializations that were gained over the same period. This suggests that "smart" specialization should be aimed at preserving (some) existing specializations in addition to gaining new ones. Our regressions indicate that the relative roles of related and unrelated variety for explaining loss of specialization are similar to the case of specialization gains. Finally, we also show that unrelated variety is also important in indicators that are derived from density, such as the Economic Complexity Outlook Index.

# 1. Introduction

The evolution of specialization patterns is now often analyzed using the related concepts of product space and smart specialization (e.g., Hidalgo et al., 2007; Balland et al. 2019). This is a data-oriented approach that starts from the idea that the dynamics of specialization depend primarily on relatedness of the activities in which specialization takes place. Thus, if we are analyzing, for example, export specialization, the idea is that there is relatedness between the products that a country already exports and some products that can be exported, and that this leads to path dependency in what countries or regions can export (Hidalgo et al., 2007). A country is more likely to develop new exports in products that are related to its current exports, rather than in products that are unrelated. The approach is also applied to production instead of trade (e.g., Xiao et al., 2019), technological specialization, often using patent data, and to regions as well as countries (e.g., Balland et al., 2019).

Relatedness is conceptualized through the metaphor of a product space, in which some products locate close to each other, and others are far away from each other (Hausmann and Klinger, 2006; 2007). The evolution of specialization is seen as the dynamic process of occupying parts of this space (by a country or a region), and this pre-dominantly takes place by either expansion of the occupied part of product space, or a move in the product space, in both cases most likely to take place in local neighborhoods of the existing specialization pattern.

Balland et al. (2019) describe the "framework for smart specialization", which they consider similar to the 'complexity' framework proposed by Hausmann et al. (2014), as a combination of two dimensions: product relatedness and product complexity. Relatedness is conceived as an "index of the relative ease with which a region might be able to develop [a specialization]" (Balland et al. 2019, p. 1259). Complexity is seen as a measure of the pay-off of new specializations, with products with higher complexity giving higher benefits. This leads to clear policy advise about smart specialization: policymakers should focus on products that are related to the existing specialization structure of the country/region, and which have high complexity.[1]

There are a number of interrelated prime measures that are used to operationalize product space. Perhaps the most important one is density (Hausmann and Klinger, 2006), which is a product-by-country measure that indicates the likelihood that a country has or will develop a specialization in a product. The other measures relevant to smart specialization are product complexity, which is a product-level measure that captures the 'sophistication' of a product, and the economic complexity index (ECI), which is a country-level measure reflecting the competitiveness of a country (Hidalgo, 2021). In a nutshell, a country can increase its ECI by covering more complex areas of product space which are within its reach as indicated by the density metric.

In this paper, we focus on the relatedness dimension of the smart specialization framework, leaving the dimension of complexity largely aside. Our main aim is to investigate the nature of the density measure in more detail. Density is supposed to capture the effects of the *specific* locality of the country in product space (i.e., how the *specific* current specializations of a country are related to possible new specializations). We hypothesize that the measure may also capture a more general effect of diversity (which is the number of products that a country specializes in) and/or ubiquity (which is the number of countries that specialize in a product) rather than the specificity of the products that it specializes in. In other words, we investigate whether density can be a measure of unrelated variety in addition to related variety.

---

[1] We will refer to countries and products throughout the paper, because our application is to international trade, but the principles of what we discuss also refer to technologies (patents), production or employment, and to regions as well as countries.

Our starting point is the literature that tests the hypothesis (in alternative ways) that countries are more likely to gain specializations in products for which they have high density. Hausman and Klinger (2007) used OLS regressions with a comparative advantage indicator as the dependent variable and density as an explanatory variable. The significantly positive sign of the density variable is taken as support for the product space idea of path dependence in specialization patterns.

The regression approach has also been followed in subsequent literature, e.g., Boschma and Capone (2015, 2016), Bahar et al. (2017), all performed similar regressions on a sample of countries and products. Alonso and Martín (2019) apply the regression approach to Mexican regions. Guo and He (2015) apply the same regression approach to industrial employment data (including production for the domestic market). These contributions generally confirm the relation between the relatedness indicators and the gain of specializations, overwhelmingly supporting the theoretical idea of product space and the path dependence of specializations that it predicts.

Coniglio et al. (2021), on the other hand, employ a non-parametric approach by comparing the distributions of density for cases where comparative advantage was actually gained and cases where comparative advantage could potentially be gained (i.e., comparative advantage did not exist). While most of the regression approaches use a "pooled" sample of countries (or regions) and products (or industries), Coniglio et al. (2021) also implement their test for individual countries, showing that some countries fit the product space idea, but others much less so.

Differently than (though complementarily to) this literature, we use a logit regression approach, both over a pooled sample of countries and on a per-country basis. Our motivation is not to assess the general predictive power of the density measure, but to decompose the metric analytically into components that reflect purely-related and unrelated variety, to understand what lies under the statistical significance of the metric in predicting the dynamics of diversification over the product space, as shown by the literature mentioned above.

We will explore the roles of related and unrelated variety space in loosing specializations (i.e., what Farinha et al., 2019 call 'exits') as well as the gain of specialization. While much (although not all) of the literature focuses on gains only, we show below that it is not always the case that the specializations that are lost have lower complexity than the ones that are gained. Therefore, there is a distinction between moving in product space or expanding occupation in product space. If a country expands its occupation of product space, it only gains specialization, and does not lose any. However, in empirical reality, countries *move* through product space, because they gain *and* loose specializations. As we will show, moving in product space is not always in the direction of, on average, higher complexity areas.

The rest of our paper is organized as follows. The next section summarizes some salient features of our data. In Section 3, we recapitulate the definition of density, and outline how density is correlated to diversity (the number of products a country is specialized in) and ubiquity (the number of countries that are specialized in a product). We illustrate these correlations by empirical data on export specialization of countries. In Section 4, we apply a regression framework to try to predict the gains and losses of specializations in our dataset. In these regressions, we distinguish between the part of density that can be associated with related variety, and the part that is not (we take the latter as unrelated variety). We focus on evaluating the predictive power of these regressions, as well as on the relative roles of related and unrelated variety. Section 5 summarizes and concludes our argument.

## 2. Setting the scene: stylized facts in our data

We use a database on the $-value of exports in 2012 and 2018, at the 4-digit Harmonized System (HS-2012) level to construct matrix *X*. At this level of aggregation, there are 1,224 products.[2] The database contains 155 countries, including all major exporters in the world. The data come from the UN's COMTRADE database, and were retrieved from the World Bank's WITS system. The most basic indicator that we calculate using these data is revealed comparative advantage (RCA), denoted by the matrix *X*, with elements

$$x_{ij} = 1 \text{ if } \frac{E_{ij}/E_j}{E_i/E} \geq 1 \text{ and } x_{ij} = 0 \text{ otherwise,}$$

where $E_{ij}$ denotes the value of exports of product *i* by country *j*, and the absence of a subscript indicates summation over the relevant dimension. The matrix *X* has dimensions *m* x *n*, where *m* is the number of products, and *n* is the number of countries, and typically, $m \gg n$. We assume that each country exports at least one product, and each product is exported by at least one country (this ensures that all elements of *X* are well-defined).

The 155 countries together gained 6,297 specializations over the period 2012 – 2018, and they lost 5,752.[3] On average, a country gained 41 specializations, and lost 37, resulting in a net gain of 4. As examples of individual countries, China gained 67 and lost 48, the US gained 31 and lost 97.

In order to highlight the importance of loss of RCA, we briefly look at complexity in combination with lost and gained specializations.[4] Figure 1 shows the average complexity of the gained specializations vs the lost specializations for the 155 countries. The bubble size indicates the economic complexity index (ECI) of the country in 2012. The solid blue bubbles indicate positive values of ECI, with larger bubbles corresponding to larger values. The empty (white) bubbles indicate negative values of ECI, with larger bubbles indicating smaller values. Thus, the large solid blue bubbles indicate the 'advanced' countries, and the large white bubbles indicate countries lagging far behind.

We observe that a large number of observations are above the 45 degrees line, especially for countries with large (i.e., positive) ECI. This means that for many (advanced) countries, complexity of lost specializations is higher than complexity of gained specializations. This clearly shows that if we are interested in evaluating how the complexity of a country's occupation of product space changes over time, we must look at losses of specializations as well as gains of specializations. Consequently, a smart specialization policy strategy must be aimed at maintaining certain specializations as well as gaining them. For our analysis, this means that we must analyze the role of related and unrelated variety for both gains and losses of specialization.

The basic premise of the smart specialization literature is that the gain of comparative advantage is path dependent: which specializations are gained depends on which specializations already exist. We investigate this idea in the next sections, but here we start with a simpler perspective. We define the basic variables diversity and ubiquity (Hidalgo et al., 2006). Diversity is a characteristic of countries, and it is equal to the number of products for which the country has

---

[2] We also ran the entire analysis in the paper using 6-digit product classes, of which there are 5,197. This does not change the conclusions in a qualitative way. These results are available on request.
[3] A gained specialization is defined as RCA equal to zero in 2012 and one in 2018, loss of specialization is the reverse.
[4] We use the standard measures for product complexity and the economic complexity index for countries, as in Hidalgo (2021).

RCA = 1. Ubiquity is a product characteristic, and it is equal to the number of countries that have RCA = 1 for the product. We measure these variables using the 2012 data.

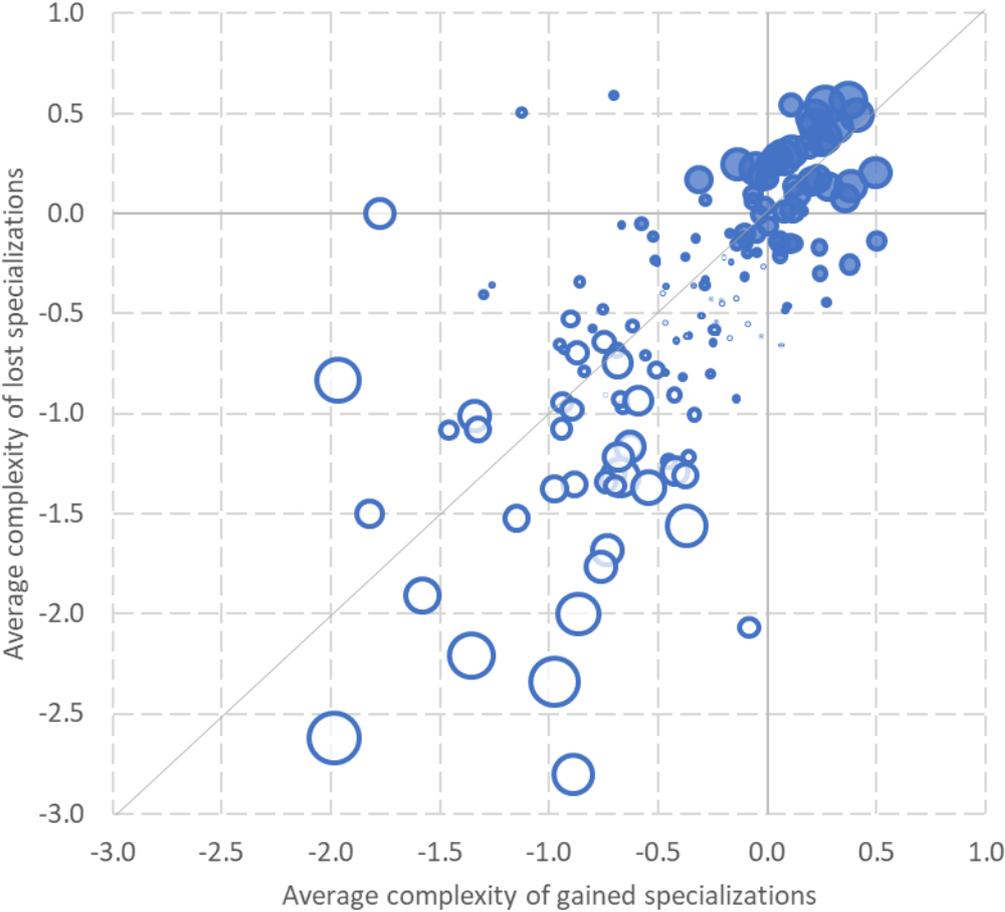

**Figure 1. Average complexity of gained and lost specializations**

We adopt a country perspective throughout the paper, i.e., we analyze gains and losses of RCA per country. Thus, Figure 2 plots the number and probability of gains and losses of RCA for each country, using the diversity of the country on the horizontal axis. The average probability of gains and losses differs widely. Gain of RCA is only possible when a country has RCA = 0 in 2012, and there are 165,775 such observations in the database, with 6,297 gains realized, which implies that the average probability for a gain is about 3.8%. On the other hand, a loss of RCA can only occur if a country has RCA = 1 in 2012, which is the case for 23,945 observations. 5,752 losses occur, which is an average probability of about 24.0%.

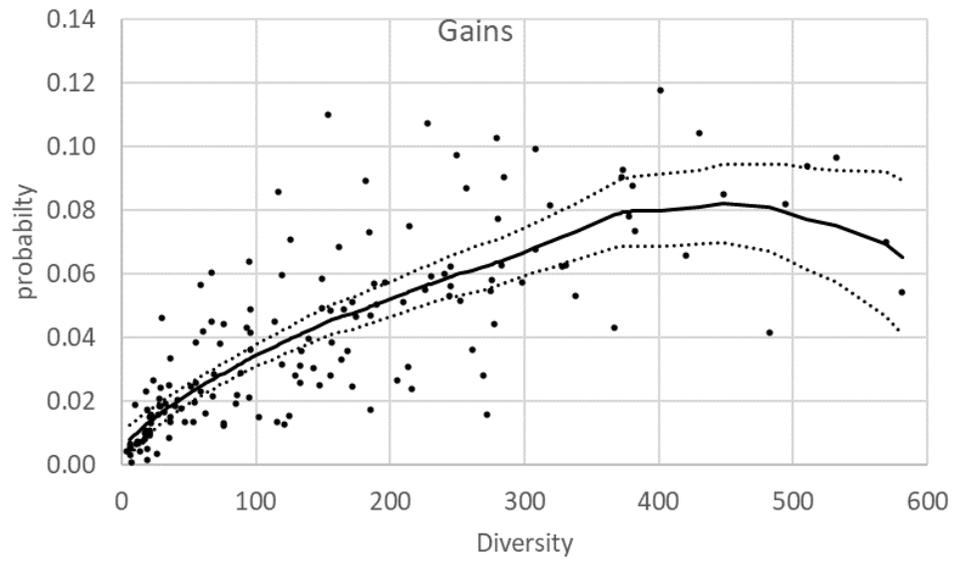
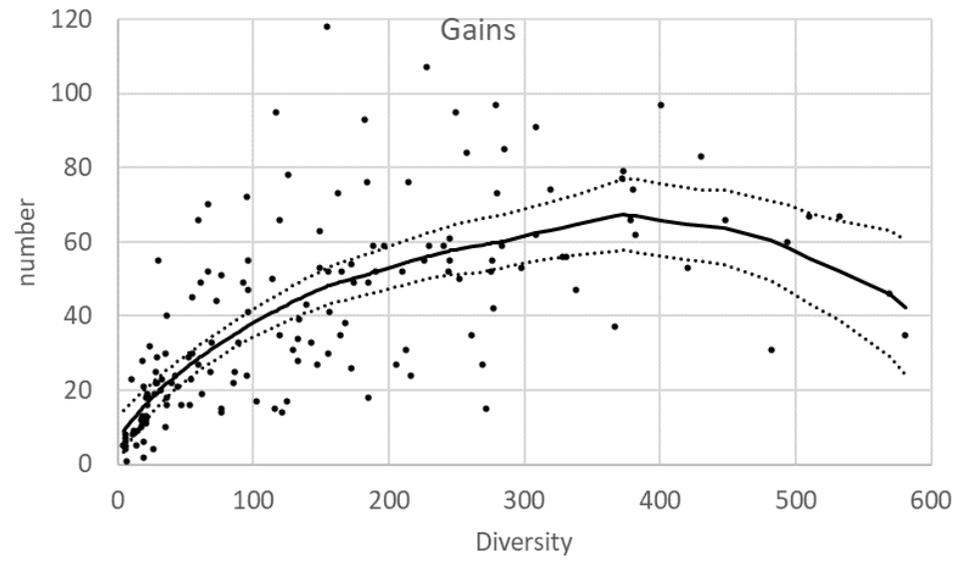
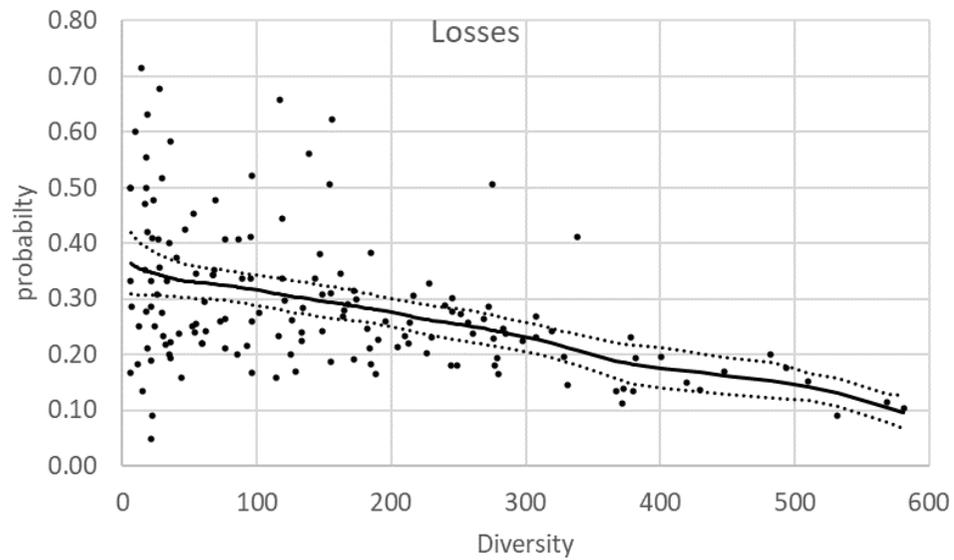
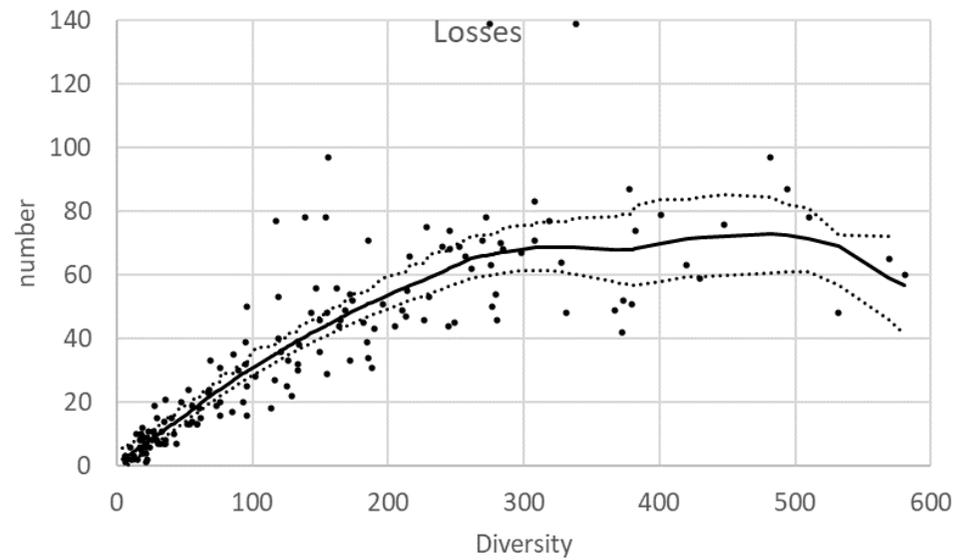

**Figure 2. Probability and number of gains and losses of specialization against diversity, observations and smoothed trends**

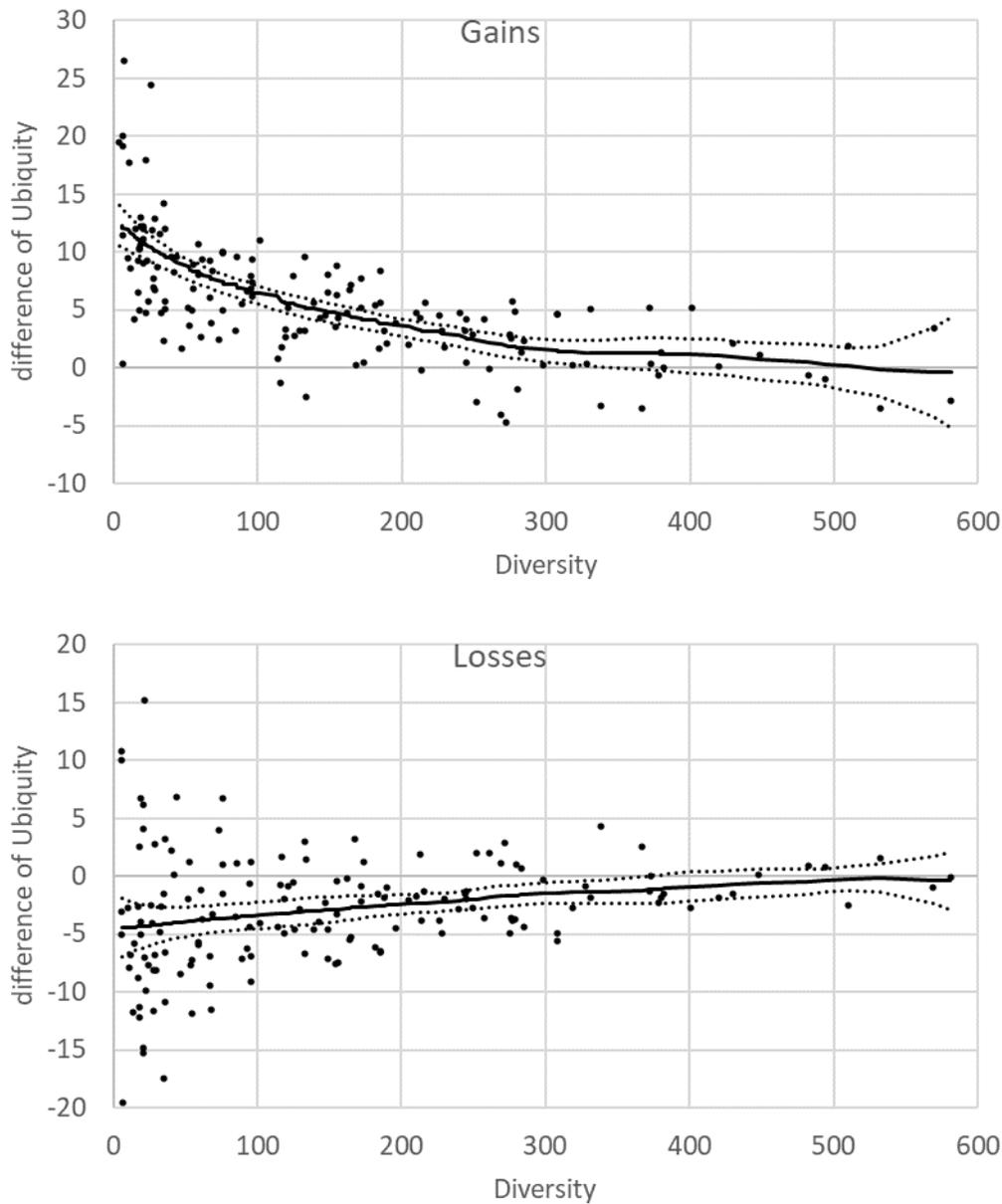

**Figure 3. Difference of ubiquity between realized and non-realized gains/losses of specialization, against diversity, observations and smoothed trends**

The dots in Figure 2 are actual observations, and the solid lines are fitted trendlines. We used a local polynomial smoothing algorithm to generate these lines. The dotted lines are 95% confidence intervals around the smoother.[5] These plots suggest that both the probability and number of gains of specialization depends positively on the diversity of the country. The probability of loss of RCA depends negatively on diversity, but the number of losses varies positively with diversity. In addition to these trends, there is some heteroskedasticity (perhaps more for losses than for gains), i.e., the variance of observations also varies along the horizontal axis, although not always in a monotonous way.

---

[5] The degree of the polynomial smooth is 1, the bandwidth for the smoother is 50, and the bandwidth for the standard error is 75. These settings are used for all polynomial smooths in this paper. In all graphs where the smoother is used, the dotted lines indicate 95% confidence intervals.

From the point of view that smart specialization is in related variety, this is a paradoxical finding, because there is nothing in the diversity measure that indicates anything about the relatedness of the products that are gained or lost, to the products that the country already is specialized in. Like diversity, ubiquity bears no obvious relatedness content, which is why we plot ubiquity in Figure 3. Because we want to keep diversity on the horizontal axis, we plot the difference between average ubiquity of the realizations (gains or losses) and the non-realizations (non-gains and non-losses, respectively) on the vertical axis. Otherwise, Figure 3 has the same layout as Figure 2. We observe, in addition to again a heteroskedastic pattern, that for low levels of diversity, there are notable differences in ubiquity between the realizations and the non-realizations, and the difference is positive for gains (realized gains have higher ubiquity than non-realized gains) and negative for losses (realized losses have lower ubiquity than non-realized gains). This difference goes to zero for higher values of diversity. This suggests that ubiquity is also related to the gain and loss of RCA.

## 3. Density and its relation to related and unrelated variety

We will now explore the nature of the density indicator (Klinger and Hausman, 2007). It is based on the idea that a country's current specialization structure (partly) determines which products are likely targets for developing new comparative advantages (smart specialization). It draws on the notion of conditional probability in RCA, computed on the basis of observed co-occurrences in $X$. The usual interpretation is that if conditional probability is high (low), the products are likely to share a high (low) degree of capabilities needed to export them with comparative advantage.

To formalize this, let $k_{pq}$ denote the number of countries that have comparative advantage both in product $q$ and in product $p$, and $s_p$ denote the number of countries that have comparative advantage in product $p$ ($s_p$ is what is usually called the 'ubiquity' of a product). Then $c_{qp} = k_{qp}/s_p$ denotes the probability that a country has comparative advantage in product $q$, conditional on the country having comparative advantage in product $p$. In matrix notation, these conditional probabilities are given by

$$C = S^{-1}XX^T,$$

where the superscript $T$ indicates a transposed matrix, and $S$ is the matrix with the corresponding row-sums of $X$ on the main diagonal and zeros elsewhere. Note that the diagonal of $S$ thus contains the ubiquity of respective products, i.e.,

$$s_{ij} = \begin{cases} \sum_{\forall k \in \{1,2,..n\}} x_{ik} & if\ i = j \\ 0 & otherwise \end{cases}$$

The density metric can be written in matrix format as follows:

$$D = \frac{C^{min}X}{C^{min}O}$$

with $C^{min} = \min(C, C^T)$, $O$ a matrix of identical dimensions as $X$ filled with only 1s, and the division being element-by-element. $C^{min}$ is the symmetrized conditional probability matrix, so that the conditional probability of $p$ on $q$ is equal to that of $q$ on $p$, using the smallest of the two elements ($c_{pq}$ and $c_{qp}$) from $C$. In this way, the asymmetric information in pair-wise conditional probabilities is transformed into a symmetric metric of product 'proximity'.

The $D$ matrix transforms the pair-wise product proximity information into a normalized (in [0,1]) metric of the overall proximity between a given product and all products that are in the current specialization portfolio of a given country. The elements $d_{ij}$ of matrix $D$ are the density of product $i$ for country $j$ ($D$ has dimensions $m \times n$, like $X$), and they reflect the sum of the cells of the product $j$ row of $C^{min}$ for which $i$ has a comparative advantage, relative to the sum of all elements in the $j^{th}$ row of $C^{min}$. Thus, a hypothetical country without any comparative advantages (a country that does not export) would have $d_{ij} = 0$ for all products $j$, and a hypothetical country that has comparative advantage in all products would have $d_{ij} = 1$ for all products $j$.

The issue that we want to investigate revolves around the fact that density is correlated both to diversity and to ubiquity (Hidalgo, 2021 notices the correlation to diversity). To see the intuition of this correlation, we note that $d_{ij}$ (density for country $j$ of product $i$) consists of the summation of conditional probabilities over the products that $j$ has a specialization for. This means that, *ceteris paribus* the conditional probability matrix $C$, the more products $j$ is specialized in, the higher $d_{ij}$ will be. As diversity of country $j$ is simply the number of products it is specialized in, this leads to a correlation between density and diversity.

Looking at ubiquity, which is the number of countries that are specialized in a product, we can also see why a correlation with density would be present. Products with high ubiquity are commonly present in the specialization portfolio of a country, and hence they tend to co-occur often with other products. Hence the conditional probabilities in matrix $C$ tend to be high for rows or columns of products with high ubiquity.

These correlations can be illustrated using our empirical data on export specialization. Table 1 documents regressions of density on diversity and ubiquity. Because diversity and ubiquity are orthogonal in our sample of all observations[6], the $R^2$ value in the regression with both variables is equal to the sum of $R^2$ values in the regressions with just one variable. The latter are, obviously, measures of the strength of the partial correlations that we discussed above.

**Table 1. Regressions of density on diversity and ubiquity**

|  | Dependent variable density | | |
|---|---|---|---|
| Constant | -0.036*** | -0.0098*** | 0.104*** |
| Diversity | 0.0009*** | 0.0009*** |  |
| Ubiquity | 0.0014*** |  | 0.0014*** |
| $R^2$ | 0.9266 | 0.9164 | 0.0102 |
| Number of observations | 189,720 | 189,720 | 189,720 |

\*\*\* denotes significance with *p*-values < 0.01.

Both variables are highly significant in all three regressions. The correlation between density and diversity appears to be rather strong and significant, and while density and ubiquity are not as strongly correlated (R ≈ 0.1), the latter is still a significant correlation. Also note that, given the orthogonality of diversity and ubiquity, the $R^2$ = 0.0102 in the regression with only ubiquity can be seen as a 0.0102 / (1 – 0.9164) = 0.1225 share of the variance that is left unexplained by diversity. The regression with both variables captures almost 93% of the total variance of density.

---

[6] Such orthogonality is generally not present in samples where only observations with RCA=1 or RCA=0 are considered, as is the case in our logit regressions below.

What does the correlation of density to diversity and ubiquity imply for density as a measure of related variety, and what does this imply for the prediction of gains or losses of specialization over time? As already stressed above, both diversity and ubiquity bear no direct relation to relatedness, i.e., these measures do not tell us anything about how products are related to each other. Hence they cannot tell us which products should be the target of a smart specialization strategy that is aimed at diversifying into products that are related to the current specialization pattern of a country. With countries as the focal point, policy advise could be to focus on products with low (or high) ubiquity, but this advice would not be based on relatedness. Similarly, we could formulate a policy advice for countries with low (or high) diversity, but again, such an advice would not be based on relatedness. Still, it is obvious from the mere correlations that are implied by Table 1 that any advice based on density contains an important part that is related to ubiquity and to diversity.

There are also indicators that are derived from density. For example, the so-called Economic Complexity Outlook Index (ECOI) is presented as a predictor of economic growth by Hausmann et al. (2014), and is a multiplicative compound indicator based on RCA, density and product complexity. Thus, if density is correlated to unrelated variety, the same may well hold for such derived indicators. In the appendix, we analyze the correlations between diversity and ubiquity on the one hand, and the ECOI on the other hand. In the remainder of the main text, we focus on trying to disentangle the role of diversity and ubiquity in predicting gains and losses of RCA from the role of relatedness *pur sang*.

## 4. Predicting specialization gains and losses

The logit regression framework is a suitable option for evaluating the potential of density for predicting gains or losses of RCA. It estimates the probability of a gain or loss happening as

$$P_{ij} = \frac{1}{1 + e^{-\beta Q_{ij}}}$$

where $P_{ij}$ is the probability of the event (gaining or losing RCA) for country *j* and product *i*, $Q_{ij}$ is a vector of observations on a range of explanatory variables (including a constant), and $\beta$ is a parameter vector that is estimated. If a constant is the only variable included in the regression, then the estimate of $P_{ij}$ will be identical across the sample of all products and countries and equal to the observed probability (3.8% for gains, 24.0% for losses). The set of variables may include variables that are constant within a country (diversity), or constant within a product (ubiquity), or vary in both dimensions (density).

Because the probabilities are a non-linear function of the explanatory variable(s), it is not possible to decompose the estimated probability into parts that are associated with the individual variables. However, such a decomposition is possible for the so-called log-odds ratio (*LOR*), which is defined as

$$LOR_{ij} = \ln\left(\frac{P_{ij}}{1 - P_{ij}}\right) = \beta Q_{ij}$$

We also note that the regression in the first column of Table 1 allows us to decompose density as follows:

$$d_{ij} = c + \delta q_j + \upsilon s_i + r_{ij}$$

where $q_j$ is diversity of country $j$, $s_i$ is ubiquity of product $i$, $r_{ij}$ is the residual from the regression that explains density as a function of diversity and ubiquity (Table 1), and $c$, $v$ and $\delta$ are the estimated parameters in that regression. The part $c + \delta q_j + v s_i$ in this equation is the predicted value of density, which is independent of the relatedness that the density indicator wants to capture. Although this part is also not strictly interpretable as unrelated variety (it is neither necessarily related, nor strictly unrelated), for simplicity we will refer to it as *unrelated* variety. The residual of the regression will be referred to as *related* variety.[7]

We start by estimating a logit model where density is the only explanatory variable (along with a constant). We denote the parameter estimate for density in this regression by $\beta_1$, and the estimated constant as $\beta_0$. Then the log odds ratio can be written as follows:

$$LOR_{1ij} = (\beta_0 + \beta_1 c) + \beta_1 \delta q_j + \beta_1 v s_i + \beta_1 r_{ij}$$

In addition to this, we also estimate a model in which diversity, ubiquity and the residual $r_{ij}$ are all used as independent variables. This will yield an alternative estimate for the probability, and for the log odds ratio:

$$LOR_{2ij} = \beta_2 + \beta_3 q_j + \beta_4 s_i + \beta_5 r_{ij}$$

Obviously, this is a more general model than the one with only density. Only if the full set of restrictions $\beta_2 = \beta_0 + \beta_1 c$, $\beta_3 = \beta_1 \delta$, $\beta_4 = \beta_1 v$ and $\beta_5 = \beta_1$ holds are the two models equivalent. Of these restrictions, the last one is the easiest to interpret: it states that the estimated coefficient on the residual in the regression with all independent variables is identical to the estimated coefficient on density in the density-only regression.

After estimating these logit models, we calculate the log odds ratio for each observation $i$, $j$, and then group these predictions by country. For each country, we then calculate the average log odds ratio for the observations that were observed to be "successful" (actual gains or actual losses), as well as for the observations that were not "successful" (non-realized gains or non-realized losses). Then, for an individual country $j$, we calculate

$$B_{1j} = \overline{LOR}_{1j}^+ - \overline{LOR}_{1j}^- = \beta_1 v (\bar{s}_j^+ - \bar{s}_j^-) + \beta_1 (\bar{r}_j^+ - \bar{r}_j^-)$$

$$B_{2j} = \overline{LOR}_{2j}^+ - \overline{LOR}_{2j}^- = \beta_4 (\bar{s}_j^+ - \bar{s}_j^-) + \beta_5 (\bar{r}_j^+ - \bar{r}_j^-)$$

where $B_{kj}$ ($k = 1,2$: either density as the only explanatory variable, or the three-variables set) is the success-*LOR* bonus for country $j$, $\overline{LOR}_{kj}^+$ and $\overline{LOR}_{kj}^-$ are, respectively, the average *LOR* (of the $k$ variety) for country $j$'s successful observations and failure observations; $\bar{s}_j^+$ and $\bar{s}_j^-$ are the average ubiquity for the successful and failure products of country $j$, again respectively; and $\bar{r}_j^+$ and $\bar{r}_j^-$ are the average residual for the successful and failure observations, respectively. The part $\beta_1 v (\bar{s}_j^+ - \bar{s}_j^-)$ or $\beta_4 (\bar{s}_j^+ - \bar{s}_j^-)$ of the success-bonus illustrates the impact of unrelated variety, while the part $\beta_1 (\bar{r}_j^+ - \bar{r}_j^-)$ or $\beta_5 (\bar{r}_j^+ - \bar{r}_j^-)$ represents unrelated variety. Note that because we calculate the success-bonus per country, diversity drops out, and unrelated variety is represented by ubiquity alone.

The estimation results for the logit regressions are documented in Table 2, for both gains of RCA and for losses. In the regression with density alone, this variable is highly significant in both cases. The sign is positive for gains, which indicates that products with higher density are more likely to be gained. Both for losses and for gains, the estimated coefficient for the residual is larger in absolute value than the coefficient on density, indicating that the restriction $\beta_5 = \beta_1$ is generally violated. For losses, the sign of density is negative, meaning that products with high density are less likely to be lost.

---

[7] Hidalgo (2021, p. 98) proposes alternative approaches to filter out the correlation between density and diversity (or ubiquity), i.e., taking z-scores of density or "dividing density by the diversity and ubiquity of a location and activity". Our approach of taking the residual of the regression in Table 1 is very similar to this.

**Table 2. Logit regressions for predicting gains and losses of specialization**

|  | Gains | | Losses | |
|---|---|---|---|---|
| Constant | -3.934*** | -4.793*** | -0.2788*** | 0.6194*** |
| Density | 5.068*** |  | -3.114*** |  |
| Diversity |  | 0.0044*** |  | -0.0021*** |
| Ubiquity |  | 0.0439*** |  | -0.0298*** |
| Residual |  | 15.38*** |  | -10.69*** |
| Number of observations | 165,775 | 165,775 | 23,945 | 23,945 |
| Pseudo-$R^2$ | 0.0473 | 0.0728 | 0.0316 | 0.0492 |

*** denotes significance with *p*-values < 0.01.

The pseudo-$R^2$ statistic is, however, low in both cases. This statistic indicates the increase in the likelihood compared to a model with only a constant. For losses, there is only a 4.7% increase if we include density in the regression, for losses, it is 3.2%. The predictive power of the logit regressions with all three variables is somewhat higher than in the one with only density, as indicated by the pseudo-$R^2$ statistics. Both for losses and for gains, a simple likelihood ratio test prefers the model with three independent variables, i.e., the restrictions that would make the two models identical are rejected as a null hypothesis. For both gains and losses of RCA, all three variables are significant, with a positive sign for gains and negative sign for losses. This means that diversified countries have a higher probability to gain RCA, irrespective of which product; that products with high ubiquity offer a higher probability of gaining RCA (irrespective of the country); and that the residual indicating related variety has a positive impact on the probability of gaining RCA. The negative signs for losses imply that all these effects apply to the probability of maintaining RCA as well (a lower probability of RCA loss).

**Table 3. Alternative indicators for predictive power of the logit regressions**

|  | Gains | | Losses | |
|---|---|---|---|---|
| Independent variables in logit | Density | Diversity, ubiquity, residual | Density | Diversity, ubiquity, residual |
| Correct positives as % of the top *LOR* | 10.2% | 13.6% | 35.3% | 38.3% |
| Correct negatives as % of the bottom *LOR* | 96.5% | 96.6% | 79.5% | 80.5% |

There is also an alternative way of looking at the predictive power of the logit regression. We can rank the observations on the log odds ratio, and pick the top 3.8% for gains (24.0% for losses).[8] As the log odds ratio and the predicted probability are monotonously related, these are the observations with the highest predicted probability. Then we can check how many of those top observations are actual gains (losses). These are the "correct positives". The observations among the top 3.8% (24.0% for losses) are the "false positives", while the actual gains not ranked (losses) in the top 3.8% (24.0%) are "false negatives", and the remaining observations are "correct negatives".

Table 3 shows the relative occurrence of correct positives and correct negatives in the logit regressions. If we only use density, there are only 10.2% correct positives in the regression for gains, and 35.3% in the regression for losses. These are relatively low numbers, and just as the pseudo-$R^2$ statistics, this shows that the explanatory power of density in explaining either gains or losses is not very large. The numbers go up slightly, especially for correct positives, if we use three explanatory variables.

Next, we look at the success bonus and its components corresponding to unrelated variety (ubiquity) and related variety (the residual). We plot the success-bonus and its components against country diversity, and again apply polynomial smoothing. The smoothed trends are documented in Figure 4. We do not plot the actual observations to make the figures more legible. Also, we do not document the smoothed curve for the residual, but instead only the ones for the total bonus and the part corresponding to unrelated variety/ubiquity. Consequently, the surface under the red curve (ubiquity) measures the contribution of unrelated variety to the success-bonus, while the part between the black curve and the red curve measures the part corresponding to related variety (the residual). The top row in the figure shows results based on the logit model with only density ($B_{1j}$), the bottom row results based on all three variables ($B_{2j}$).

For gains, we see that the red curve is above the black curve for low levels of diversity, for both logit models (only density or three variables). This means that the contribution of the residual (related variety) to the success-bonus for gains is negative for this part of the graph. Other than this, the general patterns for gains and losses are similar. The red curve (ubiquity/unrelated variety) declines with diversity, and the part corresponding to the residual generally widens. This means that the impact of related variety is relatively low for countries which are not very diversified yet, but large for diversified countries, both for gains and losses of RCA. And the median value for diversity in our sample of 155 countries is 205, while 89% of all countries have diversity lower than 300.

The bottom row of the figure looks very similar to the top row, but the ubiquity curves are somewhat higher up (relatively).[9] Thus, if we let related and unrelated variety free in the logit estimates, as opposed to restricting them as the density metric does, unrelated variety gains importance in explaining either gains or losses of RCA. This is also evident from the fact that the parameter estimates in Tables 1 and 2 show that for both gains and losses, $\beta_4 \gg \beta_1 v$ and $\beta_5 \gg \beta_1$, but the effect is relatively stronger for ubiquity than it is for the residual.

---

[8] Obviously, the 3.8 and 24.0 percentages correspond to the actually realizations. These values balance the number of false positives and false negatives equally.

[9] Note also that the vertical axis in the bottom row of the figure generally shows higher values than in the corresponding graph in the top row. This means that the success-bonus is generally larger in the model with three independent variables, because this model has higher predictive power. However, the relatively large difference here, translates to small differences in estimated probability (because the transformation from *LOR* to probability is highly non-linear).

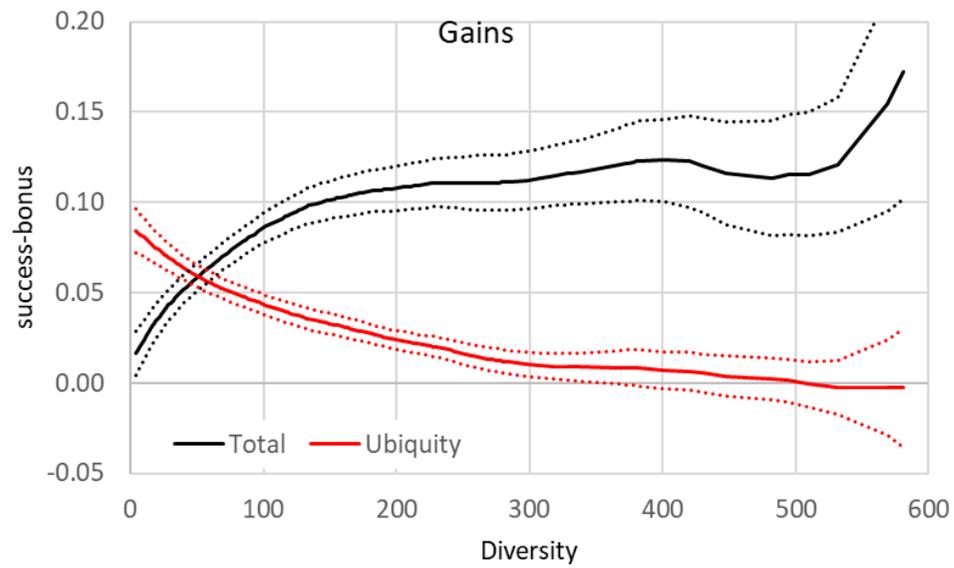
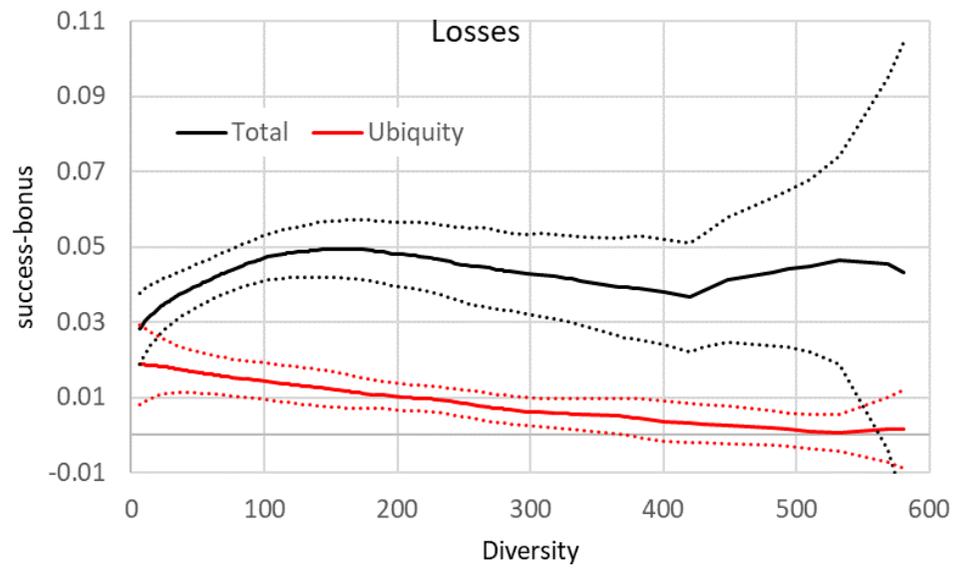
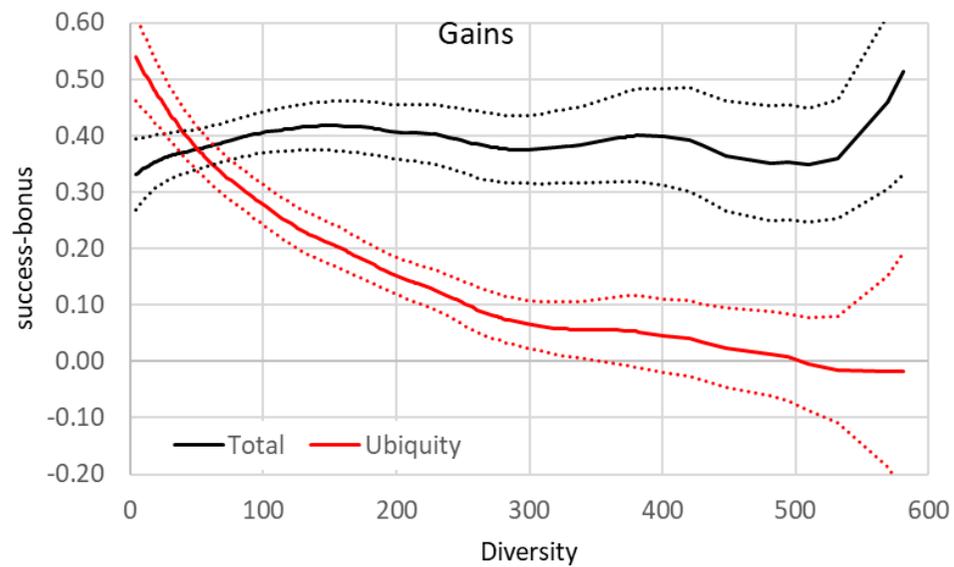
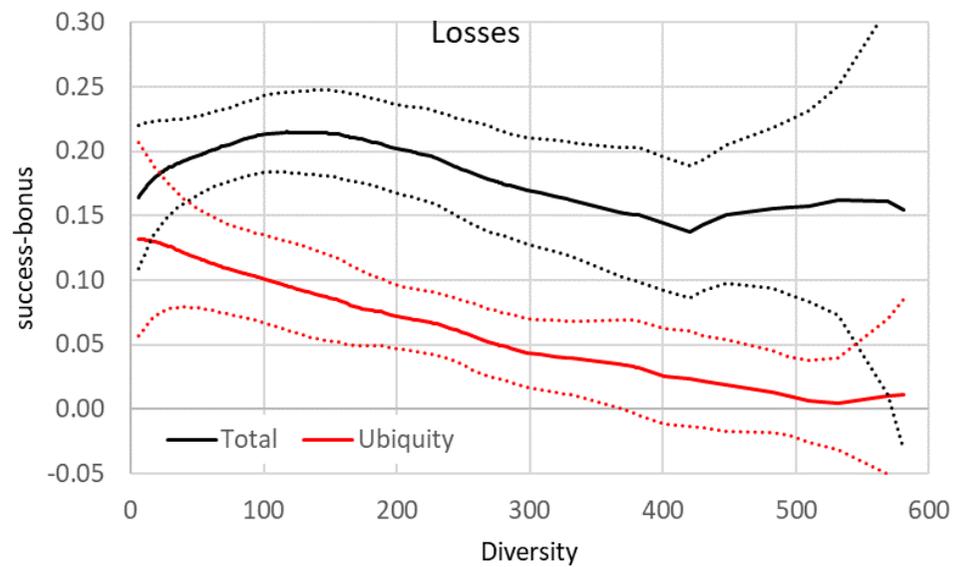

**Figure 4. Relative contribution of ubiquity and the residual to the success-bonus; model with only density on top row, model with three variables bottom row**

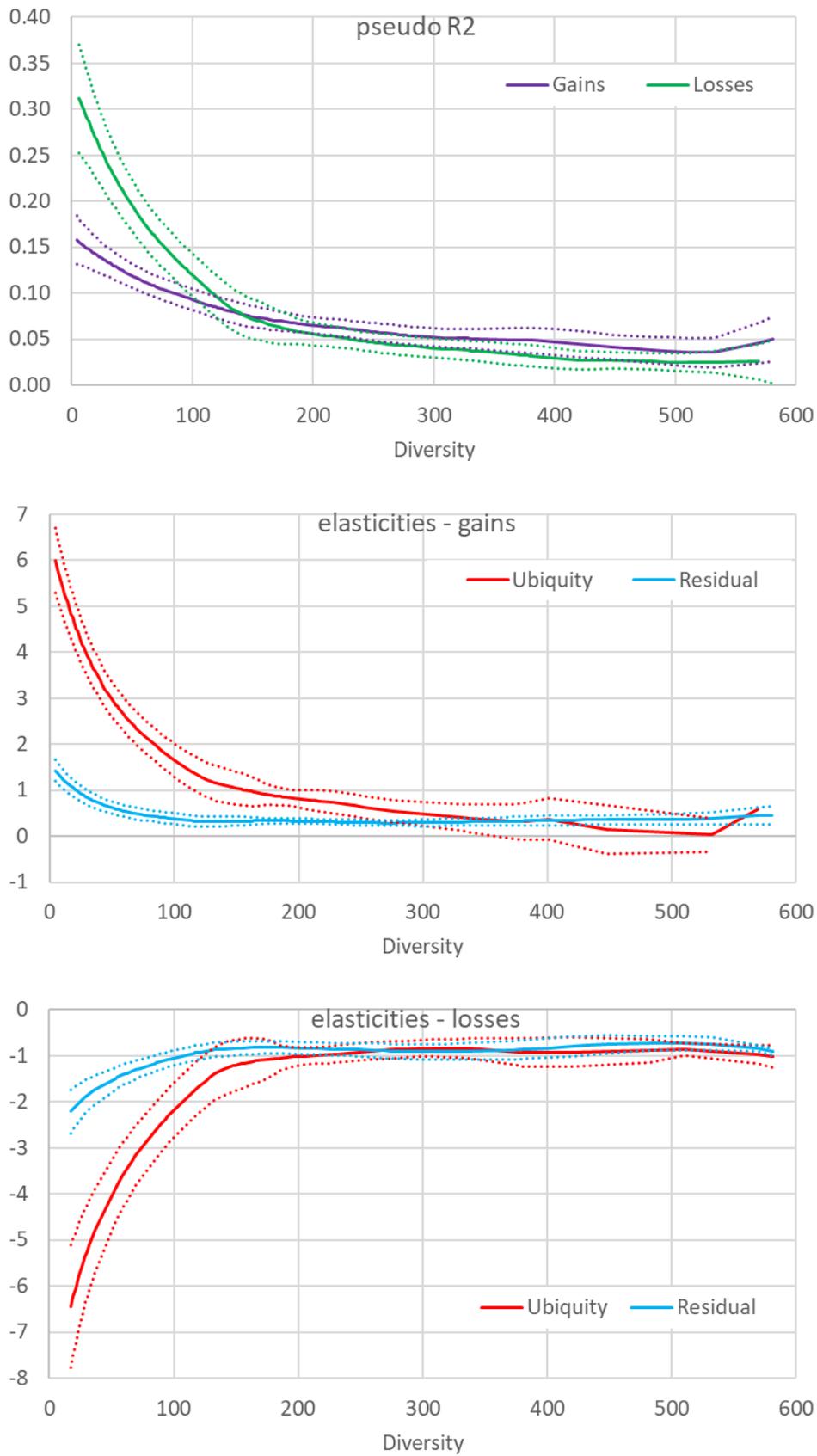

**Figure 5. Results from logit regressions for individual countries**

These results are confirmed by logit regressions at the country level. We estimate one regression for losses and one for gains for each country separately, with ubiquity and the residual as independent variables. The top panel of Figure 5 documents the pseudo-$R^2$ values of these regressions, against diversity, and smoothed as in the other figures. We generally see pseudo-$R^2$ declining with diversity, with only the largest values of diversity going below the pseudo-$R^2$ values that were observed in Table 2. The line for losses shows higher values on the left, but declines steeper. It is more difficult to estimate the dynamics of specialization (losses and gains) for countries that are already diversified than for countries at low levels of diversity.

The two bottom panels show the elasticities of the estimated probability with respect to each of the two explanatory variables. These are evaluated at the observed values for each observation, and then averaged for the country.[10] Figure 5 documents the smoothed values for these elasticities against diversity. We see that for a large range of low to middle-range values of diversity, the elasticity for ubiquity has a higher absolute value than the one for gains. This confirms that for countries at relatively low levels of diversity, an increase in ubiquity has a larger impact than in increase in relatedness, both for gains and for losses. For losses, the two lines for the marginal effects converge to very similar values, while for gains, the marginal effect for the residual eventually dominates over ubiquity, although with the confidence intervals.

## 5. Conclusions

We have argued that the density measure that is often used in the smart specialization literature captures more than product relatedness. Specifically, our results for countries' specialization in international trade show that diversity (the level of overall diversification of a country) and ubiquity (how many countries are specialized in a product) capture about 93% of the variance in density in our sample. Because the main policy recommendation from this literature is that countries (or regions) should diversify into activities that are related to their current set of activities, and because diversity and ubiquity have nothing to say about such relatedness, this is a potentially problematic finding. As an aside to our main story, our appendix shows that the importance of unrelated variety also extends to indicators that are derived from density, such as the Economic Complexity Outlook Index (ECOI).

We also argued that losses of specialization are important and deserve policy attention. In our data, the specializations that were lost by a country often represented higher product complexity than the specializations that were gained over the same period. This suggests that "smart" specialization should be aimed at preserving (some) existing specializations in addition to gaining new ones. For this reason, we put equal emphasis on gains and losses of specialization in our analysis, and in the appendix, we introduce an alternative version of ECOI that accounts for potential gains and losses of specialization.

We use a regression approach to decompose density into a part that is related variety, and a part that is unrelated variety, and we use these parts in logit regressions for the gain or loss of specialization. Although such changes in specialization are generally hard to predict (the logit regressions do not have large explanatory power), the variables for related and unrelated variety are generally significant. Deeper analysis of these regression results shows that the importance of related variety relative to unrelated variety increases with diversity of a country. In other words, the smart specialization policy advice is mostly relevant for countries that are already fairly highly diversified. Countries that show relatively low diversification levels tend to diversify into products with high ubiquity, i.e., products that many other countries are already specialized in. Because most countries have relatively low levels of diversity, a "smart" specialization strategy may therefore, in many cases, be one that attempts to diversify in a general manner, rather than focus strongly on related diversification.

---

[10] When the average value of the residual was negative, we multiplied the elasticity by minus one. This ensures that the sign of the elasticity is the same as the sign of the estimated coefficient on the residual in the logit regression.

Our regressions for the loss of specialization generally show that the same variables (related and unrelated variety) that correlate to specialization gains also correlate to specialization losses, but with opposite signs. Thus, specializations are less often lost in related variety, as well as in products with high ubiquity, and countries with high diversity tend to loose specializations less often. Our finding on relative importance of related and unrelated variety for specialization gains also holds for specialization losses: the importance of related (unrelated) variety increases (declines) with the level of diversification of a country. Thus, our policy conclusion that especially for countries that are not highly diversified yet, smart specialization can be based on unrelated variety, holds across the board for gaining as well as maintaining specializations.

Our results are based on a dataset of international trade for countries. The smart specialization literature also addresses regions, or even cities, and other activities than trade. We hope to see (and/or do) future research that addresses the relative contribution of related and unrelated variety, as well as specialization gains and losses, also in these other contexts.

**Appendix – The Economic Complexity Outlook Index and unrelated variety**

In this appendix, we analyze the ways in which the Economic Complexity Outlook Index (ECOI) is related to diversity and ubiquity, i.e., to unrelated variety. Because the ECOI is a derived indicator in which density, RCA and product complexity are included, it is supposed to capture the idea of related variety, and especially its impact on (potential) economic growth.

As will become evident from the formal exposition below, ECOI can be seen as an indicator that measures the potential complexity gains associated with related variety, in other words, the complexity that a country may gain by diversifying into products that are related to its current specializations, but in which it is not specialized yet. Density represents the relatedness information in this reasoning behind ECOI. Because our results suggest that density is also to an important extent correlated to unrelated variety, ECOI likely also captures unrelated variety.

Such a link to unrelated variety puts the interpretation of the ECOI indicator in a different light. The main text of our paper concludes that related variety may play a smaller role in explaining gains or losses of specializations that is commonly suggested. The analysis of ECOI in this appendix makes a similar and complementary point about density-derived indicators.

As before, let $x_{ij}$ denote the binary RCA of country $j$ in product $i$, let $d_{ij}$ denote the density of country $j$ in product $i$, and let $\Gamma_i$ denote the complexity of product $i$. Then the ECOI of country $j$ is defined as

$$\text{ECOI}_j = \sum_i d_{ij}(1 - x_{ij})\Gamma_i$$

The term $(1 - x_{ij})$ ensures that the summation takes place over products that the country has RCA = 0, and density captures the likelihood that the country will gain RCA in product $i$. Then ECOI is the density-weighted sum of complexity of products that the country can gain specialization in, or an expected gain of complexity. It is seen as a predictor of growth because products with high complexity are seen as having high growth potential (Hausmann et al., 2014).

In the spirit of the emphasis of loss of specialization (exit) in our main text, one may also define a complementary ECOI version, which we denote by $\widetilde{\text{ECOI}}$:

$$\widetilde{\text{ECOI}}_j = \sum_i (1 - d_{ij})x_{ij}\Gamma_i$$

Here the inclusion of $x_{ij}$ ensures that we sum over all products that the country has RCA = 1. $(1 - d_{ij})$ reflects the likelihood of losing RCA[11], so that $\widetilde{\text{ECOI}}_j$ can be interpreted as an expected loss of complexity. We may also subtract these two versions of ECOI to obtain an expected net gain of complexity:

$$\overline{\text{ECOI}}_j = \text{ECOI}_j - \widetilde{\text{ECOI}}_j = \sum_i (d_{ij} - x_{ij})\Gamma_i$$

We calculated ECOI and $\overline{\text{ECOI}}$ and regressed them on diversity and the average ubiquity of products with or without RCA.[12] The results of these regressions are in Table A1.

---

[11] Hausmann et al., 2014 refer to the metric $(1 - d_{ij})$ as the 'distance' between a product and a country.
[12] Note that $\overline{\text{ECOI}}$ is only about 13% (negatively) correlated with ECOI, indicating that these metrics contain different information. Although beyond the scope of this paper, it can be interesting to see whether $\overline{\text{ECOI}}$ can improve economic growth projections when used as an alternative (or complement) to ECOI.

**Table A1. Regressions of the Economic Complexity Outlook Index (ECOI) on diversity and ubiquity**

| Variables | (1) ECOI | (2) ECOI | (3) ECOI | (4) $\overline{ECOI}$ | (5) $\overline{ECOI}$ | (6) $\overline{ECOI}$ |
|---|---|---|---|---|---|---|
| diversity | 0.00246*** (0.000220) | | 0.00207*** (0.000231) | -0.00630*** (0.000845) | | -0.00973*** (0.00112) |
| Ubiquity | | -0.380*** (0.0543) | -0.196*** (0.0487) | | -0.658*** (0.176) | -1.477*** (0.173) |
| Ubiquity* | | | | | 0.154*** (0.0220) | -0.0175 (0.0268) |
| Constant | 0.0224 (0.0451) | 7.570*** (1.025) | 3.773*** (0.934) | 1.589*** (0.174) | 8.807*** (3.359) | 30.42*** (3.721) |
| Observations | 155 | 155 | 155 | 155 | 155 | 155 |
| $R^2$ | 0.451 | 0.242 | 0.504 | 0.267 | 0.286 | 0.523 |

Ubiquity is average ubiquity over all products that the country has RCA = 0; Ubiquity* is average ubiquity over all products that the country has RCA = 1; Standard errors in parentheses; *** p<0.01, ** p<0.05, * p<0.1

In the regression for ECOI, we include the average ubiquity of products for which the country has RCA = 0, because these are also the products that are included in the calculation of ECOI. In the regression for $\overline{ECOI}$, we include the same variable, but also the average ubiquity of products for which the country has RCA = 1, because both types of products are included in the calculation of $\overline{ECOI}$.

In the regressions for ECOI, diversity is always significant with a positive sign, and ubiquity is always significant with a negative sign. The regression with both variables accounts for half of the variance in ECOI. This shows that ECOI is far from independent from diversity or ubiquity, or, in other words, that unrelated variety is an important component not only of density, but also of the derived indicator ECOI. Countries with higher diversity tend to have higher ECOI, and countries that do not specialize in products with high ubiquity tend to have lower ECOI.

For $\overline{ECOI}$, diversity has a negative sign, as does ubiquity of the products that the country does not specialize in. Ubiquity of the products that the country specializes in has a positive and significant sign in the regression without diversity, and is not significant when diversity is included. The $R^2$ values are roughly comparable. This shows that also the 'net gains' variant of ECOI contains an important element of unrelated variety.